\documentclass[11pt,twoside,twocolumn]{article}

\usepackage{amsfonts,amsmath,amssymb,bm,abstract,balance,graphicx,natbib}

\title{\textbf{Quantum exchange interaction of spherically symmetric plasmoids}}

\author{Maxim Dvornikov
\\
\small{Institute of Physics, University of S\~{a}o Paulo,} \\
\small{CP 66318, CEP 05315-970 S\~{a}o Paulo, SP, Brazil and} \\
\small{Pushkov Institute of Terrestrial Magnetism, Ionosphere} \\
\small{and Radiowave Propagation (IZMIRAN),} \\
\small{142190 Troitsk, Moscow Region, Russia} \\
\small{E-mail: maxim.dvornikov@usp.br}}

\date{}

\begin{document}

\twocolumn[\maketitle
\begin{onecolabstract}
We study nano-sized spherically symmetric plasma structures which are radial nonlinear oscillations of electrons in plasma. The effective interaction of these plasmoids via quantum exchange forces between ions is described. We calculate the energy of this interaction for the case of a dense plasma. The conditions when the exchange interaction is attractive are examined and it is shown that separate plasmoids can form a single object. The application of our results to the theoretical description of stable atmospheric plasma structures is considered.
\\
\end{onecolabstract}]


\section{Introduction}

The exchange interaction between charged particles plays an important role for the dynamics of quantum plasmas. For the first time the exchange interaction for the degenerate electron gas in metals was considered by~\citet{Wig34}. Since then a lot of works on this subject has been published (see, e.g., the recent review by~\citet{ShuEli11} and references therein).

The exchange effects are significant for degenerate and dense gases. The degenerate quantum plasmas were discussed by~\citet{LanLif80}. It was found that, owing to the smallness of the electron mass, the electron component of plasma should be treated using the quantum mechanics whereas ions still obey a classical dynamics. However, if the temperature of ions is much lower than that of electrons, quantum effects may be important for the ion component of plasma as well. Moreover, when one considerers a structured plasma, i.e. when the parameters of such a media are not stochastic, the behavior of charged particles may obey quantum laws.

In the present work we shall discuss the situation when the exchange interaction is significant for the ion component of plasma. Firstly, in Sec.~\ref{MODEL}, we formulate the model of the effective interaction of spherically symmetric plasmoids via the quantum exchange forces between ions. Then, in Sec.~\ref{NUMEST}, we present the numerical estimates of the magnitude of this interaction for the realistic case of an atmospheric plasma. In Sec.~\ref{APPLICATION} we consider a possible application of the obtained results to the description of stable atmospheric plasmoids. Finally, in Sec.~\ref{CONCL}, we briefly summarize our results.

\section{Model}\label{MODEL}

In this section we formulate the model for the quantum exchange
interaction of two spherically symmetric plasmoids which are
radial nonlinear plasma oscillations. We compute the effective
energy of this interaction and compare it with the typical electromagnetic energy of
each of the plasmoids.

Let us discuss a plasma composed of hot electrons and singly
ionized cold ions. We can excite a spherically symmetric
oscillation of electrons in this medium. One can parameterize this
kind of oscillation using the vector of the electric field,
\begin{equation}\label{EF}
  \mathbf{E}(\mathbf{r},t) =
  \mathbf{E}_1(\mathbf{r}) e^{-\mathrm{i}(\omega t + \varphi)} +
  \text{c.c.},
\end{equation}
where
\begin{align}\label{EFampl}
  \mathbf{E}_1(\mathbf{r}) = & \mathbf{n} E_1(r),
  \notag
  \\
  E_1(r) = & A r \exp
  \left(
    -\frac{r}{2\sigma}
  \right).
\end{align}
Here $\omega$ is the frequency of the plasma oscillation,
$\varphi$ is the possible phase shift of the oscillation,
$\mathbf{n}$ is the unit radial vector, $A$ is the amplitude of
the oscillation, and $\sigma$ is the effective size of the
plasmoid. Note that the shape of the envelope of the electric field
in Eq.~\eqref{EFampl} is often used in the parametrization of
nonlinear waves~\citep{And83}.

Owing to various classical~\citep{SkoHaa80} or
quantum~\citep{HaaShu09} nonlinearities, the plasma structure
described by Eqs.~\eqref{EF} and~\eqref{EFampl} can be stable.
However, in our analysis we do not specify what kind of physical
processes underlie the plasmoid stability. Thus we shall take $A$
and $\sigma$ in Eq.~\eqref{EFampl} as phenomenological parameters.
The magnitude of these parameters will be evaluated below.

Typically the value of the oscillation frequency $\omega$ does not
exceed the Langmuir frequency for electrons, $\omega_p$. For example, $\omega$ can be 10\% less than $\omega_p$ for a classical electron nonlinearities treated in frames of the perturbation theory~\citep{SkoHaa80}. Thus ions are only
slightly involved in the oscillation process since we suggested
that ion temperature is much lower than that of electrons.
Therefore we shall suppose that ions are practically at rest and
are embedded in the negatively charged background of electrons.
Therefore the direct Coulomb interaction of ions is quite
perfectly screened by electrons.

Now, if we consider a couple of such plasmoids, then, because of
their spherical symmetry, no interaction mediated by classical
forces is possible between these structures. However, if we
suggest that ions possess nonzero static electric dipole moments
(EDM), then instantaneously their spins will be directed
along/opposite the oscillating electric field~\eqref{EF}.
Therefore the spins of ions are structured and we may expect that
some nonzero quantum exchange force may exist between these
plasmoids.

\begin{figure}
  \centering
  \includegraphics[scale=.3]{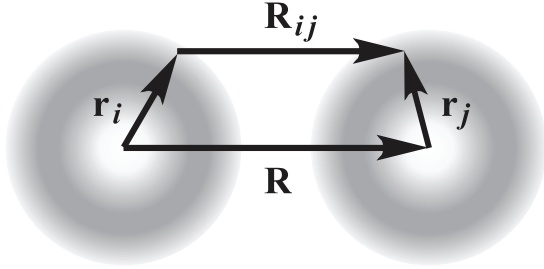}
  \caption{The schematic illustration of two interacting plasmoids separated by the distance $R$.
  The level of saturation of the grey hue corresponds to the strength of the electric field $E_1(r)$ in Eq.~\eqref{EFampl}.
  The vectors $\mathbf{r}_{i,j}$ indicate the  positions of two ions connected by the vector $\mathbf{R}_{ij}$.}
  \label{2plasmoids}
\end{figure}

For simplicity, we may discuss two identical spherically symmetric
plasma structures separated by the distance $R$ (see Fig.~\ref{2plasmoids}). The values of the
parameters $A$, $\sigma$, and $\omega$ depend on the intrinsic
properties of the surrounding media. That is why we should assume
that they are the same for both plasmoids. In other words, both
plasma structures exist owing to the same nonlinear effect. On the
contrary, the phase shift in Eq.~\eqref{EF} depends on the initial
condition of the plasmoids generation. It means that $\varphi$ may
be different for each plasmoid.
%

The majority of highly abundant neutral atoms and molecules have
integer spins. It means that a singly ionized positively charged
ion, corresponding to such a neutral particle, will have one
uncompensated electron spin. Thus we may expect that ions in our
system are half-integer-spin particles. For simplicity we suggest
that ion spins are equal to $1/2$ (see also Sec.~\ref{NUMEST}). The Hamiltonian, describing the
quantum exchange interaction of $1/2$-spin particles, is given by~\citet[p.~446]{Kit96},
\begin{equation}\label{Hamex}
  H_\mathrm{ex} = - \sum_{i \neq j}
  J_{ij} (\hat{\mathbf{S}}_i \cdot \hat{\mathbf{S}}_j),
\end{equation}
where $J_{ij}$ is the exchange integral and $\hat{\mathbf{S}}_i$
is the spin operator of an ion situated at $\mathbf{r} =
\mathbf{r}_i$. The sum in Eq.~\eqref{Hamex} is taken over the
pairs of ions belonging to different plasmoids. It means that we
do not account for the exchange interaction inside a plasmoid. Note that in Eq.~\eqref{Hamex} we keep only the spin dependent term.

When we calculate the averaging over a quantum ensemble of ions,
$\langle \dots \rangle_q$, we may suppose that the evolution of
ions belonging to different plasmoids is uncorrelated, i.e.
$\langle (\hat{\mathbf{S}}_i \cdot \hat{\mathbf{S}}_j) \rangle_q =
(\langle \hat{\mathbf{S}}_i \rangle_q \cdot \langle
\hat{\mathbf{S}}_j \rangle_q)$. This approximation is valid when
the distance between plasmoids is sufficiently great.

The mean local polarization, $\mathbf{P}=n_i \langle \mathbf{p} \rangle$ (where $n_i$ is the ion density and
$\langle \mathbf{p} \rangle$ is the mean dipole moment of ions)
of the $1/2$-spin particles gas can be calculated using the Brillouin function~\citep[p.~442]{Kit96},
\begin{align}
  B_J(x) = & \frac{2J+1}{2J}
  \coth
  \left(
    \frac{2J+1}{2J}x
  \right)
  \notag
  \\
  & -
  \frac{1}{2J}
  \coth
  \left(
    \frac{1}{2J}x
  \right),
\end{align}
where $x=p_0 E(\mathbf{r}_i,t)/T$, $p_0$ is the ion EDM, $T$ is the ion temperature, and $J=1/2$ is the total angular momentum of an ion since we suppose that ions are at the lowest energy state,
i.e. the rotational degrees of freedom do not contribute to $J$. Thus we get the ion polarization in the form,
\begin{align}\label{meanPolariz}
  \mathbf{P}(\mathbf{r}_i,t) = & g_J n_i p_0 B_{1/2}(x) \frac{\mathbf{r}_i}{r_i}
  \notag
  \\
  & =
  2 \frac{p_0^2}{T} n_i(\mathbf{r}_i,t) \mathbf{E}(\mathbf{r}_i,t),
\end{align}
where $g_J = 2$ is the $g$-factor of a $1/2$-spin ion. In Eq.~\eqref{meanPolariz} we also take $B_{1/2}(x) \approx x$ in the weak electric field limit. Finally we can find the mean local spin of ions as
\begin{equation}\label{SErel}
  \langle
    \hat{\mathbf{S}}_i
  \rangle_q =
  2 \frac{p_0}{T} \mathbf{E}(\mathbf{r}_i,t).
\end{equation}
To derive Eq.~\eqref{SErel} we take that  $\langle \hat{\mathbf{S}}_i \rangle_q = \langle \mathbf{p} \rangle_q/p_0$.

Besides the quantum ensemble averaging we should calculate the
mean value over the time interval $\tau = 2 \pi / \omega$ since
the frequency of the electric field oscillation is typically high.
Finally, using Eqs.~\eqref{Hamex} and~\eqref{SErel}, we get the
mean value of the quantum exchange Hamiltonian as
\begin{align}\label{Hamexqt}
  \langle H_\mathrm{ex} \rangle_{q\tau} = & - 8 \cos \Delta \varphi
  \left(
    \frac{p_0}{T}
  \right)^2
  \notag
  \\
  & \times
  \sum_{i \neq j}
  J_{ij} (\mathbf{n}_i \cdot \mathbf{n}_j) E_1(r_i) E_1(r_j),
\end{align}
where $\Delta \varphi$ is the phase difference in oscillations
of two plasmoids and $\mathbf{n}_i = \mathbf{r}_i/r_i$

The form of the exchange integral $J_{ij}$ in Eqs.~\eqref{Hamex}
and~\eqref{Hamexqt} depends on the ion wave functions inside a
plasmoid, which are quite difficult to calculate. Nevertheless we
may suggest that ions are in a self-consistent potential formed by
nonlinear interactions of the electron gas. In fact, this
potential provides the plasmoid stability. Therefore, according
to~\citet{LanLif91}, we may approximate the exchange integral by
\begin{equation}\label{exchin}
  J_{ij} = J_0 e^{-(\kappa_i + \kappa_j)R_{ij}},
\end{equation}
where $J_0$ is a positive constant, $R_{ij} = |\mathbf{R}_{ij}|$ is distance between
two ions (see Fig.~\ref{2plasmoids}), $\kappa_i = \sqrt{2M|\mathcal{E}_i|} / \hbar$,
$\mathcal{E}_i$ is the ion energy, and $M$ is its mass. In the
following we shall assume that ion energies are approximately
equal inside the plasmoid, i.e. $\kappa_i \approx \kappa_j \approx
\kappa = \text{const}$. It is a quite rough approximation, but it
significantly simplifies the calculations. Anyway we cannot
account for the coordinate dependence of ion energies until the
type of a nonlinearity, providing the plasmoid stability, is
specified.

Moving to the continuous density distribution and supposing that
the ion density is approximately constant and uniform,
$n_i(\mathbf{r},t) = n_0$, on the basis of Eqs.~\eqref{Hamexqt}
and~\eqref{exchin} we get for the effective potential of the
exchange interaction, $V_\mathrm{ex} = \langle H_\mathrm{ex}
\rangle_{q\tau}$, the following expression:
\begin{align}\label{Vexint}
  V_\mathrm{ex} \approx & - 8 J_0 \cos \Delta \varphi
  \left(
    \frac{p_0 n_0}{T}
  \right)^2
  \int \mathrm{d}^3 \mathbf{r}_1 \mathrm{d}^3 \mathbf{r}_2
  e^{-2 \kappa R_{12}}
  \notag
  \\
  & \times
  E_1(r_1) E_1(r_2)
  \frac{(\mathbf{r}_1 \cdot \mathbf{r}_2)}{r_1 r_2},
\end{align}
where $R_{12} = |\mathbf{R} - \mathbf{r}_1 + \mathbf{r}_2|$.


The sixfold integral in Eq.~\eqref{Vexint} can be calculated. It
is convenient to compare the final result for the effective
potential with the total electromagnetic energy of a plasmoid,
$W_\mathrm{em} = \tfrac{1}{8\pi} \textstyle{\int \langle
\mathbf{E}^2 \rangle_\tau \mathrm{d}^3 \mathbf{r}}$ (we remind
that the magnetic field is identically equal to zero for a
spherically symmetric system of charged particles), as
\begin{align}\label{VexWem}
  \frac{V_\mathrm{ex}}{W_\mathrm{em}} = & - 6.0 \times 10^3 \times
  J_0 \sigma^3 \cos \Delta \varphi
  \notag
  \\
  & \times
  \left(
    \frac{p_0 n_0}{T}
  \right)^2
  F(a,b),
\end{align}
where
\begin{align}\label{Fab}
  F(a,b) = & b \exp(-a^2/4)
  \notag
  \\
  & \times
  \big\{
    (1 + 4 b^2)
    [\mathrm{erfcx}(2b - a/2)
    \notag
    \\
    & -
    \mathrm{erfcx}(2b + a/2)]/a
    \notag
    \\
    & -
    b [\mathrm{erfcx}(2b - a/2)
    \notag
    \\
    & +
    \mathrm{erfcx}(2b + a/2)]
  \big\}.
\end{align}
Here $\mathrm{erfcx}(x) = \exp(x^2) \mathrm{erfc}(x)$ is a scaled
complementary error function as well as $a = R/\sigma$ and $b =
\kappa \sigma$ are the dimensionless parameters. We remind that $\sigma$ is the effective plasmoid size defined in Sec.~\ref{MODEL}, cf. Eq.~\eqref{EFampl}. In the next section we evaluate the magnitude of $V_\mathrm{ex}$ in a realistic situation.

\section{Numerical estimates}\label{NUMEST}

In this section we evaluate the quantum exchange interaction
of spherically symmetric plasmoids for the case of a
realistic atmospheric plasma. In particular we find the conditions when this interaction is attractive.

First, we notice that the magnitude of the effective
interaction~\eqref{VexWem} depends on EDM of an ion. The most abundant molecules in the Earth's atmosphere, like $\mathrm{N}_2$ and $\mathrm{O}_2$, are nonpolar. However, a water
molecule has a rather big EDM~\citep{Clo73},
$|p_0| = 6.18 \times 10^{-30}\thinspace\text{C}\cdot\text{m}$.
Note that the total spins of stable $^{1}\mathrm{H}$ and $^{16}\mathrm{O}$ atoms, which a water molecule is made of, are integers.
Thus, we can
consider a plasma composed of singly ionized water ions, which should be half-integer-spin particles.
We shall
suppose that they are in the lowest energy state and have spins
equal to $1/2$, for the results of Sec.~\ref{MODEL} to be valid.

The main uncertainty of Eq.~\eqref{VexWem} is in the parameters of
the exchange integral: $\kappa$ and $J_0$. To evaluate $\kappa$ we
notice that it depends on the binding energy of an ion inside a
plasmoid, $\mathcal{E}$. However, for a plasmoid to be stable
against thermal fluctuations, $|\mathcal{E}|$ should be
greater than the typical thermal energy of an ion. For a water ion
the average thermal energy is equal to $3T$ since it is a
triatomic molecule. Thus we may take that
$\kappa \sim 10^{11}\thinspace\text{m}^{-1}$
for $T \approx
300\thinspace\text{K}$ (the typical air temperature) and
$M \sim 10^{-26}\thinspace\text{kg}$
(the mass of the
water molecule). To make a rough estimate one may also assume that
$J_0 \sim 3T$.

\begin{figure}
  \centering
  \includegraphics[scale=.8]{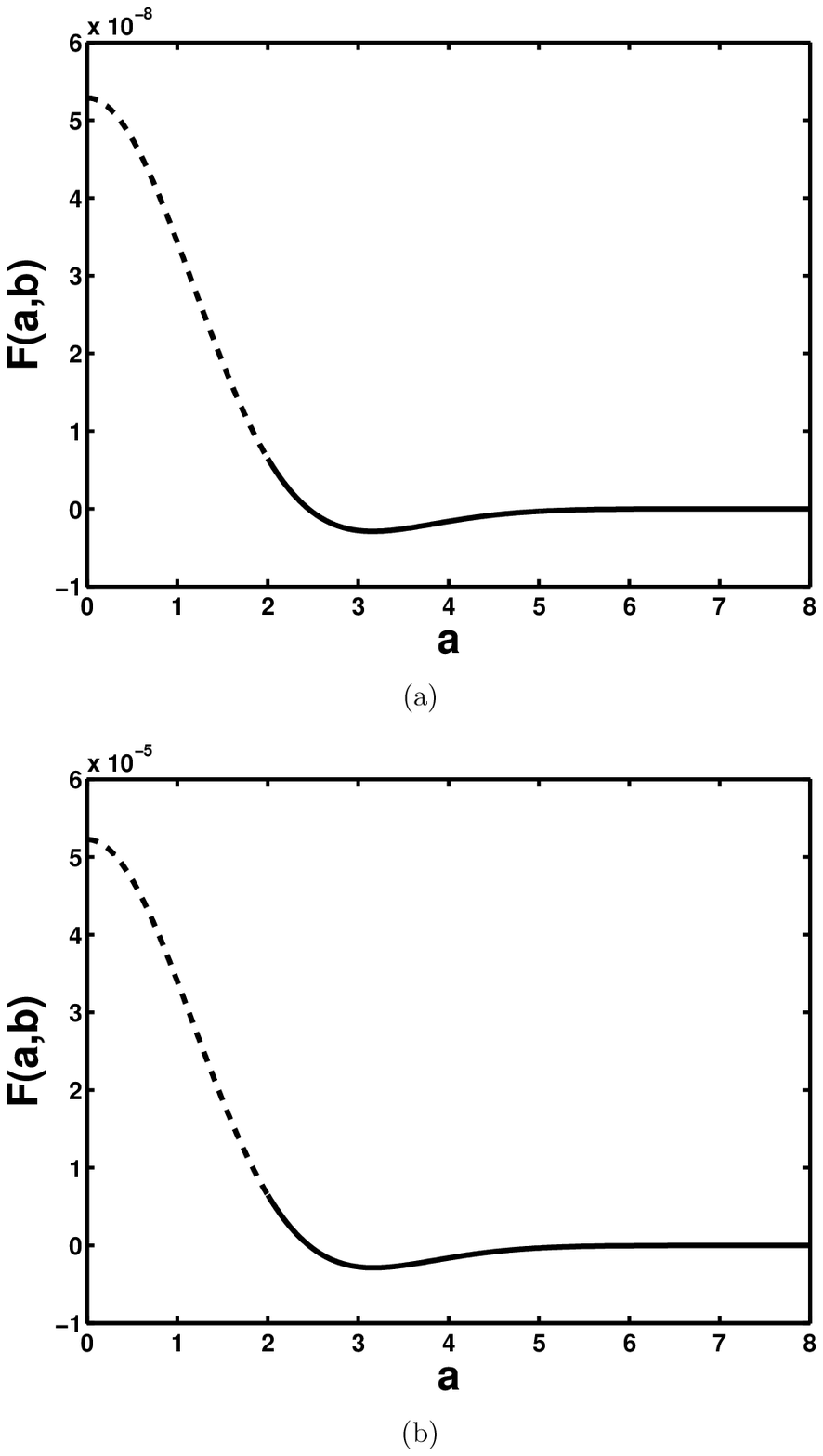}
  \caption{\label{Fab10010}
  The function $F(a,b)$ versus $a$ for different values of $b$: (a) corresponds to $b = 100$ and (b) to $b = 10$.
  The part of the curves for $a < 2$ is shown by a dashed line since it corresponds to an
  unphysical situation of almost overlapping plasmoids with $R < 2\sigma$.}
\end{figure}

\citet{DvoDvo06,Dvo12} found that stable
quantum plasma structures of the nano-scaled size
$(10^{-8}-10^{-9})\thinspace\text{m}$
can exist in the
atmosphere. Let us discuss a pair of such plasmoids with
$\sigma \sim 10^{-9}\thinspace\text{m}$.
Taking into account our estimate
for $\kappa$ we get that $b \approx 100$ for these plasmoids. The
function $F(a,b)$, given in Eq.~\eqref{Fab}, versus $a$ for $b = 100$ is shown in
Fig.~\ref{Fab10010}(a).

One can see in Fig.~\ref{Fab10010}(a) that $F(a,b) \to 0$ at $a
\gg 1$, i.e. the interaction between plasmoids vanishes at great
distance between them, $R \gg \sigma$, as it should be. Moreover,
the function $F(a,b)$ is negative at $R \gg \sigma$. Using
Eq.~\eqref{VexWem}, in this case, we get that the exchange
interaction can be attractive if $\pi / 2 < \Delta \varphi < 3 \pi
/ 2$, i.e. the oscillations in plasmoids should be in antiphase.
This result can be understood on the basis of general
Eq.~\eqref{Hamex}. Indeed, let us discuss the ball segments of
plasmoids which are in front of each other. It is clear that ions
which are in these segments will give the greatest contribution to
the effective interaction since there is a shortest distance
between them, cf. Eq.~\eqref{exchin}. If we discuss the case of
the antiphase oscillations, their spins will be mainly
parallel at any moment of time. Thus $\langle (\hat{\mathbf{S}}_i
\cdot \hat{\mathbf{S}}_j) \rangle_q > 0$ and, according to
Eq.~\eqref{Hamex}, this situation should correspond to the
effective attraction. At $b = 100$ the minimum of the function
$F_\mathrm{min} \approx - 2.9 \times 10^{-9}$ is reached at $a
\approx 3.16$.

The strength of the effective attraction between plasmoids quadratically depends on the number density of water ions. Let us suppose that we study plasma structures in the air having 100\% relative humidity at $T = 300\thinspace\text{K}$. In this case $n_0 \approx 2 \times 10^{23}\ \mathrm{m}^{-3}$. Using
Eq.~\eqref{VexWem} for these plasma parameters and supposing that $\cos \Delta \varphi = -1$ we get that
$|V_\mathrm{ex}|/W_\mathrm{em} \sim 10^{-14}$, i.e. the energy of
the quantum exchange energy is very small compared to the total electromagnetic energy of a plasmoid. However, if we assume that $n_0 \sim 10^{29}\ \mathrm{m}^{-3}$, that corresponds to the number density of molecules in liquid water, we get that $|V_\mathrm{ex}|/W_\mathrm{em} \sim 10^{-2}$. It means that quantum exchange interaction can reach a few percent of the electromagnetic interaction.

One can also notice in Fig.~\ref{Fab10010}(a) that, decreasing the
distance between plasmoids, the function $F(a,b)$ becomes
positive.  Formally the situation when $a = 0$ corresponds to
completely overlapped plasmoids. In this case, provided that the
oscillations are in phase, the spins of ions will be always
parallel. Again using Eq.~\eqref{Hamex} with $|\Delta \varphi| <
\pi/2$ we get that one has an effective attraction between
plasmoids. However we cannot discuss small values of $a<2$
since in this case there will be a strong correlation between
spins of different plasmoids which will distort the dynamics of
the system making our analysis invalid. Nevertheless, at $2 < a <
a_0$, where $a_0 \approx 2.45$ is the root of the function $F(a,b =
100)$, our results are likely to be valid. Physically this
situation corresponds to slightly touching plasma structures
without a significant overlapping.
For the values used above,
$T = 300\thinspace\text{K}$ and
$n_0 \sim 10^{23}\thinspace\text{m}^{-3}$,
we again obtain that
$|V_\mathrm{ex}|/W_\mathrm{em} \sim 10^{-14}$, provided that $\cos
\Delta \varphi = 1$. If we increase the water ions density up to $\sim 10^{29}\thinspace\text{m}^{-3}$, we get that
$|V_\mathrm{ex}|/W_\mathrm{em} \sim 10^{-2}$.

We obtained that, for realistic conditions, corresponding to a
typical atmospheric plasma which contains saturated water vapor, the energy of the quantum exchange
attraction is only a small fraction of the total electromagnetic energy
of a spherical plasma structure. Thus the exchange
interaction cannot significantly influence the dynamics of
oscillations inside a plasmoid. The exchange interaction can reach a few percent of the electromagnetic interaction if we study plasmoids in denser media. Hence one may expect the coagulation of
separate nano-sized plasma structures excited, e.g., in liquid water.

We also notice that, if we lower the ion temperature, the quantum
exchange interaction should reveal itself more intensively. The
function $F(a,b)$ for $b = 10$, that corresponds to a temperature
of a few Kelvin degrees, is shown in Fig.~\ref{Fab10010}(b). One
can see that, in this case, the quantum exchange interaction is
several orders of magnitude stronger, as it should be, since
quantum effects are more essential at low temperatures. However
this case is unlikely to be implemented in an atmospheric plasma.

At the end of this section we mention that the assumption that
mean ion spins are always parallel the electric field was made in
Sec.~\ref{MODEL}, cf. Eq.~\eqref{SErel}. We remind that the
electric field inside a plasmoid is rapidly
oscillating, see Eq.~\eqref{EF}. Thus, for Eq.~\eqref{SErel} to be valid,
the typical time of the ion depolarization in an oscillating
electric field, i.e. the time of the spin-flip, $\tau_d$, should
be much shorter than $1/\omega$. The spin-flip time can be
estimated as $\tau_d \sim I/\hbar \approx 10^{-12}\ \text{s}$,
where
$I \sim 10^{-46}\ \text{kg}\cdot\text{m}^2$
is the moment of
inertia of a water ion. The oscillations
frequency for the observed plasmoids does not exceed several
MHz~\citep[p.~211]{KliKut10,BycNikDij10}. Therefore Eq.~\eqref{SErel} is always valid.

\section{Application}\label{APPLICATION}

In this section we discuss the application of the results, obtained
in Secs.~\ref{MODEL} and~\ref{NUMEST}, to the description of stable
atmospheric plasma structures. We show that the quantum exchange interaction of plasmoids can result in the formation of a composite object.

A long-lived atmospheric plasmoid, called a ball lightning (BL),
is a luminous object, having the form of a sphere of
$(0.1 - 0.2)\thinspace\text{m}$
in diameter. Numerous trustworthy reports
on such an object, as well as its photo and video recordings,
weigh heavily to the reality of this natural phenomenon. Nonetheless the
existence of BL is still a puzzle for the modern physics because of
its extraordinary properties. The numerous theoretical models of
BL, including very exotic ones, are listed in the recent review by~\citet{BycNikDij10}.

Taking into account the fact that in the majority of cases BL
appears mainly during a thunderstorm, it is natural to suggest
that it is a plasma based phenomenon. However the reported
life-time of BL, which can be as long as several minutes, is in contradiction with the typical existence time of an unstructured atmospheric plasma, which is about several ms. \citet{DvoDvo06,Dvo12} constructed a BL model based on
nonlinear quantum oscillations of electron gas in plasma. For
quantum effects to come into play, the typical size of a plasmoid
should be tiny. The detailed analysis of~\citet{DvoDvo06,Dvo12} showed that
the typical size of a plasmoid should be about
$(10^{-8}-10^{-9})\thinspace\text{m}$
for the existence of stable
spherically symmetric quantum oscillations of electrons. Note that the BL model based on radial plasma oscillations was also discussed by~\citet{Fed99,Shm03,Ten11}.


The nano-scaled size of a plasmoid predicted
by~\citet{DvoDvo06,Dvo12} can explain that sometimes BL can easily
pass through tiny holes. Despite the recent experimental
achievements in the generation of nano-sized BL-like
structures reported by~\citet{Mit08,KliKut10,ItoCap12}, it is still difficult to reconcile the visible
dimensions of a natural BL with the results
of~\citet{DvoDvo06,Dvo12}. Nevertheless, we may suggest that the core of BL,
which still can be small, is a coagulate of separate kernels, each
of them being a spherically symmetric quantum oscillation of
electrons. The mechanism, which holds these kernels together, may
be the quantum exchange interaction between ions described in
Secs.~\ref{MODEL} and~\ref{NUMEST}.

In our opinion, separate plasma oscillation kernels are created
with arbitrarily distributed oscillations phases. Then, owing to
the quantum exchange attraction, the parts of a future BL
self-organize and form a natural plasmoid. A composite BL model is
confirmed by the observational facts that a natural plasmoid often
divides into several independent objects~\citep[p.~206]{BycNikDij10}
and that the motion of smaller parts inside BL is sometimes
visible~\citep[p.~233]{BycNikDij10}.
A complex plasmoid consisting
of several hot kernels was recently generated in a laboratory by~\citet{OreMav11}.
Note that a model of BL
consisting of separate parts was also considered by~\citet{Nik06}.

Using the estimates of Sec.~\ref{NUMEST} one finds that the quantum exchange interaction can be responsible to the coagulation of separate plasmoids only if the density of water ions is rather high, $n_0 \sim 10^{29}\ \mathrm{m}^{-3}$, which corresponds to the case of liquid water. Saturated water vapor, having $n_0 \approx 2 \times 10^{23}\ \mathrm{m}^{-3}$, is not sufficiently dense to provide the required attractive force between plasma structures. It means that the described mechanism may be important at the stage of the formation of a natural BL in case it appears inside a drop of rain water. We also mention that the results of our work may be applied for the analysis of the experiments made by~\citet{Sha02,Ver08}, who generated laboratory plasmoids in electric discharges in water.

We should mention that, for the first time, the idea that the exchange
interaction can be important for the BL stability was discussed
by~\citet{Neu37}, who suggested that electrons in a dense and
relatively cold plasma can be held together by this kind of
interaction providing the plasma cohesion. However, as it was
later shown by~\citet{LanLif80}, accounting for the
exchange effects in quantum plasma just slightly enhances its
pressure, making impossible the plasma confinement.

The hypothesis that exchange interaction between electrons is
responsible to the cohesion of plasma of a natural plasmoid was later
revisited by~\citet{KulRum91}. Unlike~\citet{Neu37}, who studied
the exchange interaction between free electrons, \citet{KulRum91} considered
the situation when an electron in plasma is in a bound
state with an ion. In this case, provided that spins of the majority
of electrons are equally directed, cf. Eq.~\eqref{Hamex},
there can be an attractive interaction resulting in the formation
of a quantum liquid. However, it is unclear how to structure
electron spins in a quite hot plasma, discussed
by~\citet{KulRum91}, at the absence of an external field. Note that electrons can be polarized only by a magnetic field since electron's EDM has not been discovered yet~\citep{Hud11}.

In the present work we consider the effective interaction of separate nano-sized plasmoids using the concept of quantum exchange forces. However, contrary to~\citet{Neu37,KulRum91}, here we suggest that ions, rather than electrons, are involved in this kind of interaction. If an ion has a great EDM (like $\mathrm{H}_2\mathrm{O}^{+{}}$), one may expect that ion spins will be structured in an oscillating electric field of a plasmoid, cf. Eq.~\eqref{EF}. The detailed analysis of Secs.~\ref{MODEL} and~\ref{NUMEST} showes that a small attraction between plasmoids is possible if there is a certain phase shift in oscillating structures created in a dense plasma.

The magnitude of the quantum exchange interaction strongly depends on ion's EDM. Water, which is highly abundant in the Earth's atmosphere, has a molecule with a great EDM. It can explain the fact why BL appears mainly during or after a thunderstorm, when air is enriched with water vapors~\citep[p.~246]{BycNikDij10}. It is necessary to mention that the importance of EDM of charged particles to the stability of BL was previously considered by~\citet{Ber73,Sta73,Mes91,Gil03}. Contrary to the studies of~\citet{Ber73,Sta73,Mes91,Gil03}, in the present work it is suggested that the existence of a plasmoid is provided by a nonlinear interaction of electrons. The interaction of EDM of ions is accounted for to explain the coagulation of separate nano-sized plasmoids into a single object, which can be a natural BL.

At the end of this section we mention that BL observations reveal the variety of characteristics of this natural phenomenon. All the reported properties of BL does not seem to be reconciled in frames of a single model. Thus, perhaps several types of glowing objects are taken as BLs. In this work we construct a theoretical model of a low energy BL which has an internal structure. The explanation of possibly existing atmospheric plasma structures which possess a huge energy, up to several MJ~\citep[p.~230]{BycNikDij10}, cannot be explained within our model.

\section{Conclusion}\label{CONCL}

In conclusion we mention that in the present work we have discussed the effective interaction of spherically symmetric plasma structures via the quantum exchange forces. Each of the plasmoids is taken to be a nonlinear radial oscillation of electrons. In Sec.~\ref{MODEL} it has been obtained that a nonzero interaction can exist in such a system. This interaction can be attractive or repulsive depending on the phase shift in plasmoids oscillations. At big distances between plasmoids there is an effective attraction if oscillations are in antiphase.
If oscillations are in phase, an effective attraction is also possible, provided the distance between plasmoids is relatively small. However the latter case requires an additional special analysis since some unaccounted spin correlation effects can significantly influence the dynamics of the system.

In Sec.~\ref{NUMEST} we have made numerical estimates of the magnitude of the quantum exchange interaction. We have discussed a dense plasma having a significant fraction of water ions. It has been found that mutual effective attraction of plasmoids can reach a few percent of the total electromagnetic energy of a plasmoid in cases of both short and great distances between plasma structures if the density of ions is high, $n_0 \sim 10^{29}\ \mathrm{m}^{-3}$. If one studies the attraction of plasma structures existing in a medium containing saturated water vapor with $n_0 \approx 2 \times 10^{23}\ \mathrm{m}^{-3}$, the quantum exchange interaction is insufficiently strong to provide the coagulation of plasmoids. Taking into account that no other types of interaction between spherical plasmoids are allowed, we may conclude that separate plasma structures can be held together by the exchange interaction, in case they exist in a dense medium like liquid water.

In Sec.~\ref{APPLICATION} we have discussed the application of the obtained results to the construction of a theoretical model of a spherical atmospheric plasmoid, BL. \citet{DvoDvo06,Dvo12} demonstrated that a stable nano-sized plasma structure, based on nonlinear quantum oscillations of electrons, can exist in the atmosphere. In the present work we have shown that separate kernels, with independent radial plasma oscillations, can form a single structure owing to the exchange interaction of ions. This phenomenon may happen at the initial stages of the BL evolution, when plasmoids appear in a drop of rain water. Thus we have developed a composite BL model which is not excluded by the observational data.


\section*{Acknowledgments}
I am grateful to S.~I.~Dvornikov, S.~Lundeen, A.~I.~Nikitin,
E.~V.~Orlenko, and V.~B.~Semikoz for helpful discussions as well as to
FAPESP (Brazil) for a grant. A special thank is addressed to A.~V.~Nemkin for his help on the illustration preparation.

\balance

\end{document}